# Evaporation of quark drops during the cosmological quark–hadron transition

L. Rezzolla*

*SISSA, Via Beirut 2-4 – 34013 Trieste, Italy*

J.C. Miller†

*SISSA, Via Beirut 2-4 – 34013 Trieste, Italy*
*Nuclear and Astrophysics Laboratory, University of Oxford, Keble Road, Oxford OX1 3RH, England*
*Osservatorio Astronomico di Trieste, Via Tiepolo 11 – 34131 Trieste, Italy*

O. Pantano‡

*Dipartimento di Fisica "G.Galilei", Via Marzolo 8 – 35131 Padova, Italy*

(February 11, 1995)

We have carried out a study of the hydrodynamics of disconnected quark regions during the final stages of the cosmological quark-hadron transition. A set of relativistic Lagrangian equations is presented for following the evaporation of a single quark drop and results from the numerical solution of this are discussed. A self-similar solution is shown to exist and the formation of baryon number density inhomogeneities at the end of the drop contraction is discussed.

## I. INTRODUCTION

According to the standard picture of cosmology, a phase transition during which a plasma of free quarks and gluons was converted into hadrons should have occurred at about $10^{-5}$ s after the Big Bang. At present it is not completely clear whether this transition would have been first order or of a higher order [1], although recent lattice calculations favour it being first order for the realistic case of two degenerate light $u$ and $d$ quarks and a heavier $s$ quark (of up to 400 MeV) [2]. The interesting consequences that a first order transition could have produced for the early history of the universe [3,4] provide the motivation for studying this case.

The work described here is within the scenario of a first-order transition starting with the nucleation of hadronic bubbles in a slightly supercooled quark-gluon plasma. The bubbles are thermodynamically favoured and grow with the phase interface moving as a deflagration front and with the quark-gluon plasma being progressively transformed into a plasma of hadrons. The dynamical time scale of the growth is likely to be much shorter than the universe expansion time scale, and so the hadronic bubbles grow until they meet and eventually coalesce. This next stage of the transition is probably characterized by disordered motions and dissipative processes which transform the kinetic energy associated with the ordered motion of the quark regions into internal energy of the two phases. The temperature jump between the two phases becomes extremely small with both phases being almost at the critical temperature $T_c$ and the rate of transformation of the quark-gluon plasma being controlled by the overall expansion of the universe. When more than half of the space is occupied by hadronic matter, regions containing quark matter start to become disconnected and, while evaporating, they tend to become spherical under the effects of surface tension.

We concentrate here on the hydrodynamics of a single evaporating quark drop, making use of the mathematical and numerical experience developed for the previous study of growing hadronic bubbles [5–8]. A particular feature of our study is a demonstration of the existence of a self-similar solution which is likely to be relevant for some of the final stages in the shrinking away of quark drops and which provides the basis for our discussion of the final stages of the transition. These stages are particularly important since it is then that relics could be produced which might then survive until later epochs. The possible astrophysical consequences of a first order transition have been extensively investigated in the literature both with regard to the formation of baryon number inhomogeneities [9–11] and their possible effects on primordial nucleosynthesis and also for the production of primordial magnetic fields [12,4]. We will comment on these expectations and discuss the implications of the results obtained here for discussions of the formation of baryon number inhomogeneities.

The following is a brief outline of the contents of the paper. The set of relativistic hydrodynamical equations and junction conditions at the deflagration front are reviewed in Section II. The existence and derivation of the self-similar solution for a shrinking drop is discussed in Section III, while Section IV is dedicated to the presentation of the method for the computation. Here we comment on the specification of the initial conditions and discuss numerical tests which have been carried out. Results from the computations are presented in Section V, where we also discuss their consequences for the forma-tion of baryon number density inhomogeneities. Finally, conclusions and perspectives for future developments of the work are presented in Section VI. In this paper, we use units for which $c = \hbar = k = 1$ and denote standard partial derivatives with a comma.

---
*Electronic address: rezzolla@tsmi19.sissa.it
†Electronic address: miller@tsmi19.sissa.it
‡Electronic address: pantano@mvxpd5.pd.infn.it

## II. RELATIVISTIC HYDRODYNAMICAL EQUATIONS

We here consider a single, isolated, spherical quark drop, whose surface is inward-moving as a consequence of the evaporation. The drop has initial dimensions which are much smaller that the typical distance between disconnected quark regions (often taken to be of the order of $10^{14} - 10^{16}$ fm), but larger than the shortest mean free path of the particles which are not strongly-interacting. These are electromagnetically and weakly interacting particles (mainly photons, muons, electrons, neutrinos and their respective antiparticles), and although they are not directly involved in the phase transition, they can provide a long-range energy and momentum transfer at a scale comparable with their mean free path $\lambda$. For neutrinos and antineutrinos $\lambda \sim 10^{13}$ fm, while for electromagnetically interacting particles $\lambda \sim 10^4$ fm. A quark drop with initial dimensions of the order of $10^5$ fm is essentially transparent to the neutrinos (which will be neglected here), while the strongly interacting matter is in thermal and mechanical equilibrium with the electromagnetically interacting particles. Because all of these are essentially massless particles, we can consider them as components of a generalized "radiation fluid" and their dynamical and energetic contribution is taken into account by making a suitable modification of the number of degrees of freedom in the equations of state for the quark and hadron phases.

The thermal and mechanical equilibrium ceases to hold when the drop has reached dimensions comparable with the mean free path of the radiation fluid which then becomes progressively more decoupled from the strongly-interacting matter. Although a complete treatment of the evaporation of a quark drop, involving the solution of the relativistic radiative transfer problem together with the pure hydrodynamical one, is in principle possible, many of the main features of the evaporation process already emerge in a computation in which the radiation fluid is always considered as completely coupled to the strongly interacting ones. Performing a calculation in which long-range energy and momentum transfer is taken into account is certainly a more difficult task and making an error analysis without first having a clear underlying hydrodynamical solution would be much more complicated. For these reasons the set of hydrodynamical equations used here is not the one presented in [7,8], where the relativistic radiative transfer problem was solved for a growing bubble, but rather the equivalent one without radiative transfer. This was described in [5,6] to which the reader is referred for further details.

Consider a Lagrangian formulation in which the space-time metric has the spherically symmetric line element

$$ds^2 = -a^2 dt^2 + b^2 d\mu^2 + R^2 \left( d\theta^2 + \sin^2\theta \, d\varphi^2 \right), \quad (1)$$

with $\mu$ being a comoving radial coordinate having its origin at the centre of symmetry. The system of hydrodynamical equations then can be written as

$$u_{,t} = -a \left[ 4\pi R^2 \frac{\Gamma}{w} p_{,\mu} + G \left( \frac{m}{R^2} + 4\pi p R \right) \right], \quad (2)$$

$$R_{,t} = a u, \quad (3)$$

$$\frac{(\rho R^2)_{,t}}{\rho R^2} = -a \frac{u_{,\mu}}{R_{,\mu}}, \quad (4)$$

$$e_{,t} = w \rho_{,t}, \quad (5)$$

$$\frac{(a w)_{,\mu}}{a w} = \frac{e_{,\mu} - w \rho_{,\mu}}{\rho w}, \quad (6)$$

$$m_{,\mu} = 4\pi R^2 e R_{,\mu}, \quad (7)$$

$$\Gamma = 4\pi \rho R^2 R_{,\mu} = \left( 1 + u^2 - 2 G m / R \right)^{1/2}, \quad (8)$$

$$b = \left( 4\pi R^2 \rho \right)^{-1}. \quad (9)$$

Here $u$ is the radial component of the fluid four-velocity in the associated Eulerian frame, $R$ is the Schwarzschild circumference coordinate, $\Gamma$ is the general relativistic analogue of the Lorentz factor, $\rho$ is the relative compression factor, $w$ is the specific enthalpy ($w = (e + p)/\rho$, with $e$ being the energy density and $p$ the pressure). For small net baryon number, it is reasonable to take $e$ and $p$ as depending only on temperature, and then the equations of state for the two phases of the strongly interacting matter can be written as

$$e_h = (g_h + g_e) \left( \frac{\pi^2}{30} \right) T_h^4, \qquad p_h = \frac{1}{3} e_h, \quad (10)$$

$$e_q = (g_q + g_e) \left( \frac{\pi^2}{30} \right) T_q^4 + B, \quad (11a)$$

$$p_q = (g_q + g_e) \left( \frac{\pi^2}{90} \right) T_q^4 - B, \quad (11b)$$

where the hadron medium is considered as consisting of massless, point-like pions, while the quark phase is described by the *Bag Model* equation of state, with $B$ being the "Bag" constant. We here take $g_q = 37$, $g_h = 3$ and $g_e = 9$, where $g_e$ accounts for the degrees of freedom of the electromagnetically interacting particles.

As in the hadronic bubble growth computation, we treat the transition region as a discontinuity surface with the fluid and metric variables on either side being linked by suitable junction conditions. Making use of the Gauss-Codazzi equations [13], the junction conditions for energy and momentum can be expressed respectively as

$$[(e+p)ab]^{\pm} = 0, \tag{12}$$

$$[eb^2\dot{\mu}_s^2 + pa^2]^{\pm} = \frac{\sigma}{2}f^2 \left\{ \frac{1}{ab}\frac{d}{dt}\left[\frac{b^2\dot{\mu}_s}{f}\right] + \frac{f_\mu}{ab} \right.$$
$$\left. + \frac{2}{fR}(b\dot{\mu}_s u + a\Gamma) \right\}^{\pm}, \tag{13}$$

where $\dot{\mu}_s = d\mu_s/dt$ is the interface velocity ($\dot{\mu}_s < 0$), $f = (a^2 - b^2\dot{\mu}_s^2)^{1/2}$ and $\sigma$ is the surface tension. The labels $\pm$ indicate quantities immediately ahead of the transition front (quark phase) and immediately behind it (hadron phase) and $[A]^{\pm} = A^+ - A^-, \{A\}^{\pm} = A^+ + A^-$. Other supplementary junction conditions follow from continuity across the interface of the metric quantities $R$, $dR/dt$, and $ds$

$$[R]^{\pm} = 0, \tag{14}$$

$$[au + b\dot{\mu}_s \Gamma]^{\pm} = 0, \tag{15}$$

$$[a^2 - b^2\dot{\mu}_s^2]^{\pm} = 0, \tag{16}$$

and from the time evolution of the jump in the mass function $m$

$$\frac{d}{dt}[m]^{\pm} = 4\pi R^2 \left[b\dot{\mu}_s \Gamma e - apu\right]^{\pm}. \tag{17}$$

For the circumstances under consideration here, it is satisfactory to calculate the jump of the mass function at the interface as $[m]^{\pm} = 4\pi R_s^2 \sigma$. Because the drop surface moves as a deflagration front, one further equation is needed expressing the rate at which the quark matter is transformed into hadrons at the phase interface. A suitable expression is obtained by setting the hydrodynamical flux $F_H$ into the hadron region equal to the net thermal flux $F_T$ into it

$$F_H = -\frac{aw\dot{\mu}_s}{4\pi R_s^2 (a^2 - b^2\dot{\mu}_s^2)}$$
$$= \left(\frac{\alpha}{4}\right)(g_h + g_e)\left(\frac{\pi^2}{30}\right)(T_q^4 - T_h^4) = F_T. \tag{18}$$

Here $\alpha$ is an accommodation coefficient ($0 \leq \alpha \leq 1$) which can be associated with the "transparency" of the phase interface to the thermal flux and is, at least in principle, calculable from theory. Hereafter we will assume $\alpha = 1$. A discussion of the behaviour of solutions for lower values of $\alpha$ can be found in [6]. (Note that in deriving (18) we have taken the unit space-like four-vector normal to the interface as the one pointing towards the centre of the drop.)

A main requirement for the numerical solution of the system of hydrodynamical equations and junction conditions at the interface is that there should be a correct causal treatment of the deflagration front [6,7]. If the phase interface is to be treated as a discontinuity, then it seems that the only satisfactory way of accomplishing this is by using a system of characteristic equations into which the hydrodynamical equations (2)–(9) can be transformed. For this we have used the equations and methodology presented in [6], to which the reader is referred for further details.

### III. SELF-SIMILAR SOLUTIONS

In general a self-similar motion can be expected to occur when there are no intrinsic length or time scales in the system. When this is the case, the hydrodynamical equations can be re-written in terms of a single dimensionless variable obtained as a suitable combination of the parameters of the problem.

Self-similar growth of bubbles in first order cosmological transitions has been considered in the literature both in the case of detonation fronts [14–16] and deflagration fronts [17,18]. For bubble expansion, self-similar motion appears only when the bubble radius is small enough so that there is no interaction between bubbles, but large enough so that surface tension effects are negligible. In the case of drop contraction we expect that self-similar motion should set in when the drop radius is reasonably smaller than the mean distance between quark regions, so that one can neglect the interaction between drops, but large enough so that surface tension effects are negligible.

In this section, it is convenient to work with the special relativistic form of the hydrodynamical equations written in terms of Eulerian coordinates (here denoted by $r$ and $t$). Gravity does not play any important role on the scale of the drops being considered here and so it is quite sufficient to consider only the special relativistic equations for discussing these self-similar solutions. For spherical symmetry, the standard conservation and continuity equations are

$$\frac{\partial}{\partial t}\left[(e+pv^2)\gamma^2\right] + \frac{\partial}{\partial r}\left[(e+p)\gamma^2 v\right]$$
$$= -\frac{2}{r}\left[(e+p)\gamma^2 v\right] \tag{19}$$

$$\frac{\partial}{\partial t}\left[(e+p)\gamma^2 v\right] + \frac{\partial}{\partial r}\left[(ev^2+p)\gamma^2\right]$$
$$= -\frac{2}{r}\left[(e+p)\gamma^2 v^2\right] \tag{20}$$

$$\frac{\partial}{\partial t}(\rho\gamma) + \frac{\partial}{\partial r}(\rho\gamma v) = -\frac{2}{r}\rho\gamma v \tag{21}$$

where $\gamma = (1-v^2)^{-1/2}$ is the relativistic Lorentz factor and $v$ is the three velocity of the fluid. In the case of a one-parameter equation of state, such as the one which we are using, it is sufficient to solve just equations (19) and (20) and equation (21) is then needed only if one wishes to calculate $\rho$. For a relativistic system, the only possible choices for a similarity variable are $\xi = \pm r/t$ [19]. We take $\xi = r/t$ and write $e = e(\xi)$, $\rho = \rho(\xi)$ and $v = v(\xi)$. In terms of $\xi$ the system (19)–(21) becomes:

$$[(c_s^2\xi^2 - 1)v^2 + 2\xi(1-c_s^2)v + (c_s^2 - \xi^2)]\frac{dv}{d\xi}$$
$$= -\frac{2v}{\xi}c_s^2(1-v^2)(1-\xi v) \qquad (22)$$

$$\frac{c_s^2}{(e+p)}\frac{de}{d\xi} = \left(\frac{\xi - v}{1-\xi v}\right)\left(\frac{1}{1-v^2}\right)\frac{dv}{d\xi} \qquad (23)$$

$$\frac{1}{\rho}\frac{d\rho}{d\xi} = \frac{1}{(e+p)}\frac{de}{d\xi} \qquad (24)$$

where $c_s = (\partial p/\partial e)^{1/2} = (1/3)^{1/2}$ is the sound speed. The similarity variable $\xi = r/t$ can be viewed either as the position of a point in the solution at a given time, or as the velocity at which a given point in the profile of the solution is moving. This velocity is to be clearly distinguished from the fluid velocity $v$ at the point described by $\xi$. A self-similar flow which is linearly expanding with time is naturally described in terms of a positive similarity variable $\xi$. Conversely, a self-similar contracting flow, such as the one occurring for an evaporating quark drop, is described in terms of negative values of $\xi$. In this case, the time can be thought of as progressing through negative values and tending to zero as the radius of the contracting drop tends to zero.

When solving equations (22)–(24) numerically, it is useful to notice their property of invariance under the simultaneous transformations $\xi \to -\xi$ and $v \to -v$ which has the consequence that it is only necessary to solve them in one of the half-planes $\xi \in [0, \pm 1]$ in order to know the solution in the whole interval $\xi \in [-1, 1]$. In Figs. 1 and 2, we have plotted the results of numerical integrations for the functions $v(\xi)$ and $e(\xi)$. The dashed lines in Fig. 1 represent points for which the derivative $dv/d\xi$ in (22) becomes infinite, i.e. for which

$$(c_s^2\xi^2 - 1)v^2 + 2\xi(1-c_s^2)v + (c_s^2 - \xi^2) = 0, \qquad (25)$$

which has the roots

$$v = \frac{\xi \pm c_s}{1 \pm c_s\xi}. \qquad (26)$$

Expression (26), which is also the special relativistic formula for velocity composition, expresses the fact that the fluid velocity, measured relative to an observer moving at $\xi$, equals the speed of sound. A physical solution for $v$ cannot be extended across these points, since it would then be double valued, and a discontinuity must be introduced. The upper right quadrant of Fig. 1 shows the solutions of the similarity equations for an expanding system (such as a spherically expanding hadron bubble [17,6]). The upper left quadrant shows the equivalent solutions for a contracting system (such as an evaporating spherical quark drop), while the lower quadrants provide the corresponding solutions for negative values of the fluid velocity (note that these do not necessarily represent physically realistic configurations).

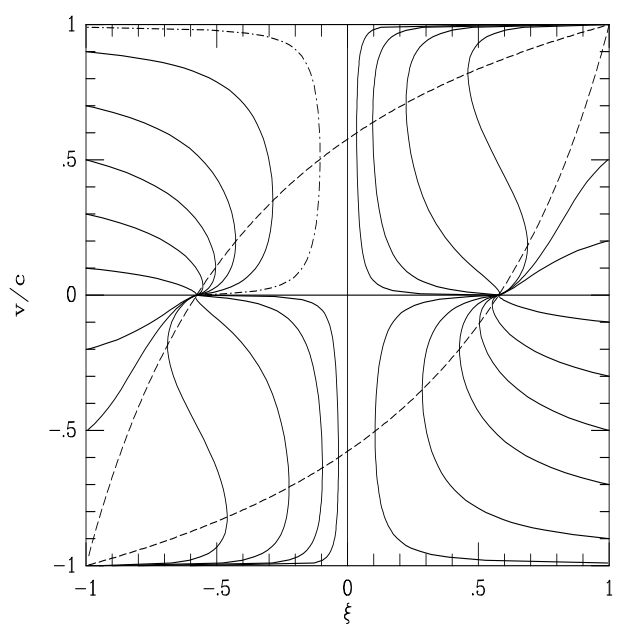

FIG. 1. Solutions of the similarity equations for the velocity $v$. The relevant quadrant for a contracting drop is the upper left-hand one. The point $\xi = -c_s$ represents the sonic radius, while the centre of the drop is at $\xi = 0$ at any given time. Parts of solution curves similar to the dot-dashed one are used in construction of the similarity solutions shown in Fig. 3, where they appear in a mirror image.

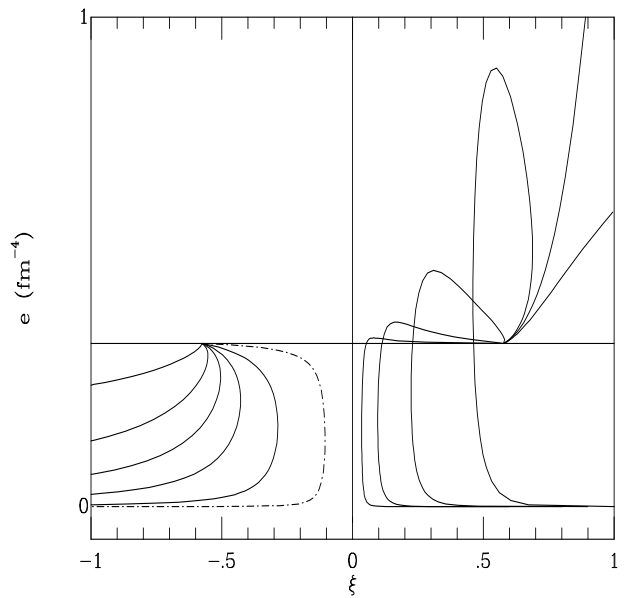

FIG. 2. Solutions of the similarity equations for the the energy density $e$. The solutions have been suitably normalized to the value of the energy density at the sonic point and correspond to solutions in the upper quadrants of Fig. 1. The relevant curves for a contracting drop are in the negative part of the $\xi$ plane. The point $\xi = -c_s$ represents the sonic radius, while the centre of the drop is at $\xi = 0$ at any given time. Parts of solution curves similar to the dot-dashed one are used in construction of the similarity solutions shown in Fig. 4, where they appear in a mirror image.

For the contracting quark drops which we are considering here, the point $\xi = 0$ represents the centre of the drop at any given time, $\xi = -c_s$ is the "sonic radius" and $\xi = -1$ is the edge of the past light cone. The only self-similar flow solution in Figs. 1 and 2 which can be taken to extend for all $\xi$ is the trivial solution $v = 0$, $e = const$. Any other solution must be the result of a "patching" of different solution curves with the joins being either via a weak discontinuity (where the function is continuous but has a derivative of some order which is not continuous), or via a full discontinuity (where the function itself is not continuous). The first applies in the case of the edge of a rarefaction or compression wave, while the second occurs in the case of shocks or combustion fronts.

The situation for a contracting drop is radically different from that for an expanding bubble. For the contracting drop, the deflagration front is moving inwards and spherical symmetry imposes that the only possible state ahead of it is one with zero velocity and constant density since the only self-similar solution satisfying the condition $v = 0$ at $\xi = 0$ is the trivial one. At the drop surface, a discontinuity corresponding to the deflagration front must be introduced in order to join onto the relevant self-similar solution curve for the flow behind it. Once $T_h$ has been specified at the interface (i.e. $T_h^-$), the junction conditions give the velocity of the front and the state of the fluid adjacent to it, which correspond to points in the $(v, \xi)$ and $(e, \xi)$ planes of Figs. 1 and 2. The solution for the medium behind the front then follows the continuous curves passing through the relevant points.

For a weak deflagration (the physically interesting case), the self-similar solution behind the front is characterized by decreasing velocity, which goes to zero at the sound radius ($\xi = -c_s$), and increasing energy density. At the sound radius, the solution then joins (via a weak discontinuity) onto another one with zero velocity and constant energy density. Note that a weak discontinuity, (which propagates at the local sound speed) can *only* be located at the point $\xi = -c_s$ and the sonic radius is then the edge of the perturbed flow region at any given time. These weak deflagration solutions are illustrated with the solid line curves in Figs. 3 and 4. Different curves are drawn for different values of the temperature in the hadron phase adjacent to the phase interface, with $\hat{T}_h^- = T_h^-/T_c$) ranging between 0.95 for the leftmost solid curve and 0.61 for the rightmost solid curve. The latter represents a Chapman-Jouguet deflagration, for which the front moves at the sound speed relative to the fluid behind [20]). This marks the transition from weak deflagrations (which are physically realistic) to strong deflagrations (which are physically forbidden and are indicated with dashed lines in Figs. 3 and 4). Note that the rightmost solid curve and the leftmost dashed curve correspond to adjacent but distinct values of $\hat{T}_h^-$.

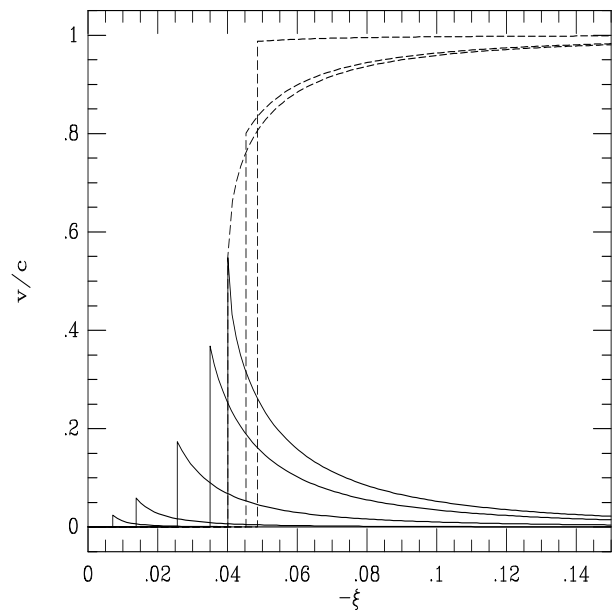

FIG. 3. Self-similar curves for the velocity for different values of the temperature in the hadron phase adjacent to the interface. The solid curves represent weak deflagration solutions with the rightmost solid curve being a Chapman-Jouguet deflagration. The dashed curves represent strong deflagration solutions beyond Chapman-Jouguet limit and are physically forbidden.

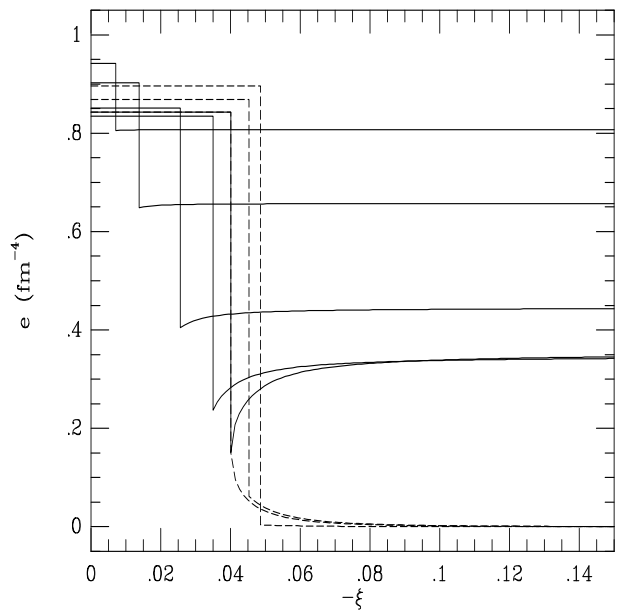

FIG. 4. Self-similar curves for the energy density for different values of the temperature in the hadron phase adjacent to the interface. The solid curves represent weak deflagration solutions with the rightmost solid curve being a Chapman-Jouguet deflagration. The dashed curves represent strong deflagration solutions beyond Chapman-Jouguet limit and are physically forbidden.

The solution for the medium behind a weak deflagration front can be seen as a compression wave since the energy density is increasing to the value in the unperturbed fluid. This occurs because a fluid element crossing the phase interface experiences a strong decompression from its original state and is subsequently compressed to the background value. Conversely, for the strong deflagration solutions, the medium behind the front is accelerated away, reaching a velocity approaching $c$ at infinity, while the energy density decreases through a rarefaction wave, going to zero at infinity. Note that as $\hat{T}_h^-$ decreases from 1, the temperature of the quark phase first decreases but then reaches a minimum (at $\hat{T}_h^- = 0.70$) and starts to increase again. This is consistent with the fact that the function $T_q = T_q(T_h)$ is not monotonic (see Fig. 3 of [21]).

When analyzing self-similar deflagration solutions for contracting drops, a number of novel features are encountered. Three main differences with respect to the equivalent solutions for expanding bubbles are:

*i*) In the case of an expanding bubble, the deflagration is always preceded by a compression wave fronted, in principle, by a shock (which has negligible amplitude in the case of a hadronic bubble nucleated with small supercooling [6,8]). For a contracting drop, the deflagration can only be preceded by a solution with zero velocity and constant energy density.

*ii*) The solution for the medium behind the deflagration front in an expanding bubble could differ from a constant state with zero velocity only if the bubble were expanding supersonically, in which case it would have to be followed by a rarefaction wave [18]. The medium behind the deflagration in a contracting drop is never at rest, but rather is outflowing with the velocity tending either to zero at the sonic point (in the case of weak deflagrations) or to the light velocity at infinity (in the case of strong deflagrations).

*iii*) Depending on the degree of supercooling, the deflagration front in an expanding bubble could move with velocities up to the sound speed (and possibly beyond) relative to the centre of symmetry. Irrespective of whether the deflagration is weak or strong, the surface of a contracting drop always moves very subsonically relative to the centre of symmetry (*i.e.* at the surface, $|\xi| \ll c_s$).

The existence and properties of the self-similar solutions for a contracting drop will be used when specifying initial conditions for the numerical solution of the full hydrodynamical equations (as described in the following Sections). The initial conditions used are similar to the leftmost solid curves in Figs. 3 and 4.

## IV. NUMERICAL STRATEGY

### A. Organization of the grid

For following the evaporation of quark-gluon drops, we have used a numerical code constructed on the basis of the strategies adopted previously for calculating the growth of hadronic bubbles [5–7]. As in the previous computations, the present code makes use of a composite numerical technique in which a standard Lagrangian finite-difference method is used to solve the hydrodynamical equations in the bulk of both phases, while a system of characteristic equations and a set of junction conditions are solved in the grid zones adjacent to the phase interface. We have used a spherically symmetric Lagrangian grid having comoving coordinate $\mu$ and with its origin at the centre of the drop. The grid has variable spacing with the width of each zone being twice that of the zone inside it (*i.e.* $\Delta\mu_{j+1/2} = \Delta\mu_{j-1/2}$), apart from the two central zones which have equal width. This arrangement allows bubble expansion or drop contraction to be followed through changes of many orders of magnitude.

A major difference with respect to the previous calculations is that we are here following an inward-moving phase interface rather than an outward-moving one. This has necessitated a different organization for the solution of the system of characteristic equations and the use of modified algorithms for the calculation of quantities in the grid zones immediately adjacent to the interface. Moreover, in order to keep constant the number of grid zones within the drop (as required for maintaining accuracy), it has been necessary to implement a "re-gridding" routine which creates a new zone inside the quark phase every time the interface crosses a zone boundary during its inward motion.

Regridding can be quite a delicate matter and, if the implementation is not a good one, there is a danger of introducing instabilities into the solution, particularly if function fitting routines are used. With our choice of grid structure, it is possible to avoid the use of any fitting algorithm. Our procedure is that every time the interface crosses a zone boundary (in the inward direction) the central zone is divided into two equal parts and all of the other zones are re-labelled (*i.e.* $j \to j+1$). This strategy maintains intact the structure of the grid (the first two central zones still have the same width, while the others are increasingly spaced) and limits the recalculation of new quantities only to the central zones. To keep constant the total number of zones (and thus avoid increases in the computational time), the outermost zone is removed every time the central zone is divided into two. Experience has shown that the loss of information involved in this does not create any problem since the outer edge of the grid is always very distant from the interface.

This re-gridding scheme has several advantages: it is extremely simple to implement, it is suitable for use during changes of many orders of magnitude in the drop radius, and it keeps constant the number of zones within the drop, hence maintaining resolution and numerical accuracy.

## B. Initial Conditions and Numerical Tests

The specification of initial conditions in a time dependent calculation can be considered as a "problem within the problem". When there is no obvious physical solution, it can be useful to extract information from the code itself by observing its response to trial initial data. This has been the case with the present calculation, in which the initial conditions (which have turned out to correspond to self-similar solutions), were originally "suggested" by the code itself.

Before initial data based on a self-similar solution was implemented, other choices of initial data were observed to relax, after some readjustments, to a solution having a flat density profile and zero velocity within the quark region, and an outward velocity profile falling off roughly as $R^{-2}$ in the hadron region. The subsequent evolution of this solution exhibited self-similar features and this then stimulated the investigation for a similarity solution. It is important to stress that it could happen that a self-similar solution which is mathematically possible might nevertheless not be realized in practice. In this respect, the weak deflagration similarity solution has turned out to be strikingly "robust". As a test, initial conditions with an inward-pointing velocity field inside the drop and zero velocity outside, were specified on the zero-time hypersurface. After some excursions, the solution converged towards the corresponding similarity one. Identical results have been obtained also in the case when the solution was started with very irregular and noisy initial conditions.

During the intermediate stages of the transition the universe has been re-heated by the latent heat released by the quark confinement, and the temperatures in both phases have become close to the critical temperature for the transition. The disordered motions resulting from the bubble percolation and coalescence tend to produce homogeneity in the temperature profile in both phases and the dynamics of the transition is then driven by the universe expansion, whose cooling effect is compensated by the latent heat released. During the last stages of the transition, the reduced quark volume fraction is no longer able to provide the amount of energy necessary to keep the increased hadron volume fraction at $T_c$. As a consequence $T_h$ is free to decrease and a small non-zero temperature jump drives the following hydrodynamical evolution of the disconnected quark regions. In view of this, we assume that all quantities have reached a high degree of homogeneity in each phase and then the situation is regulated by the temperature jump between the two phases. All of the variables, except for the fluid velocity $u$, are therefore taken to have "step-like" profiles. Note that this is consistent with the self-similar solution for small supercooling and low interface velocity (see Fig. 4).

Working within this scenario, we first specify the drop radius $R_s$ and the temperature of the hadron phase $T_h$ and the corresponding temperature in the quark phase $T_q$ is then calculated from the special relativistic form of the junction conditions (see eq. (15) in [21]). The next quantity to calculate is the value of the interface velocity $\dot{\mu}_s$ which is obtained from (18) as

$$\dot{\mu}_s = -4\pi R_s \rho_+ a_+ \frac{(1+\chi^2)^{1/2}-1}{\chi}, \qquad (27)$$

where

$$\chi = \frac{\alpha \pi^2 (g_h + g_e)}{60} \frac{(T_+^4 - T_-^4)}{(e + p + 4\pi^2 g_e T^4/90)_+}, \qquad (28)$$

and, using the freedom in the choice of reference values for $\rho$ and $a$, we have set $\rho_+ = a_+ = 1$. Rewriting the junction condition (12) as

$$\rho_- = \rho_+ \frac{(e+p)_-}{(e+p)_+} x, \qquad (29)$$

where $x = a_-/a_+$ and combining this with (16), it is possible to obtain the following equation

$$x^4 - \left[1 - \left(\frac{b_+ \dot{\mu}_s}{a_+}\right)^2\right] x^2 - \left[\frac{(e+p)_+}{(e+p)_-}\left(\frac{b_+ \dot{\mu}_s}{a_+}\right)\right]^2 = 0, \qquad (30)$$

whose solution provides the required value for $a_-$ and consequently for $\rho_-$. The last quantity to calculate is $u_-$, and the expression for this follows directly from (15) as

$$u_- = \frac{\dot{\mu}_s}{a_-}(b_+ - b_-), \qquad (31)$$

where, following the similarity solution, we have taken $u_+ = 0$ and also $u = 0$ for all points inside the quark drop. From the similarity solutions for weak deflagrations shown in Fig. 3, it can be seen that the velocity profile just behind the phase interface can be well approximated by a solenoidal flow, for which

$$uR^2 = const = u_- R_s^2. \qquad (32)$$

This approximation is an excellent one in the case of small supercooling. Expression (32) is applied out to the point where the value for $u$ given by this becomes less than that for the background universe, which is taken to follow the spatially flat Robertson–Walker solution. The expression for $u$ then follows from (8), with $\Gamma$ being taken equal to 1, and is

$$u = \left(\frac{2Gm}{R}\right)^{1/2}. \qquad (33)$$

Equations (27)–(33) provide all of the data required on the zero-time hypersurface.

For the present consideration of drop evaporation at the cosmological quark-hadron transition, it is expected that the degree of supercooling would be rather small and this case is our primary interest. However, in order to gain a good overall perspective of the situation, it is interesting to consider also what would happen if the supercooling were greater and so we have investigated values of $\hat{T}_h$ ranging between 0.999 and the Chapman-Jouguet point ($\hat{T}_h = 0.61$). We found it convenient and satisfactory to continue to specify the initial data as outlined above also in these cases.

## V. RESULTS

When studying the hydrodynamics of an evaporating quark drop it can be useful to compare this process with the evaporation of a water drop. The main difference between the two situations is that while the first is exothermic, the second is endothermic. This means that the water drop will continue to evaporate as long as there is an efficient transport of energy from the inner regions to the surface where it will be lost during the evaporation. In a similar way, it is important that during the quark drop evaporation there should be an efficient mechanism in the hadron phase for transporting the latent heat away from the phase interface. The geometry of the process and the complexity of the phenomena at the deflagration front, prevent any simple but accurate analysis being possible and only a numerical simulation can give a deeper insight into the details of the evolution. The present calculation is the first complete relativistic computation of an evaporating cosmological quark drop, although a more simplified analysis was carried out by Kajantie and Kurki-Suonio [22]. Numerical integrations of the hydrodynamical equations discussed in the previous sections over the whole permitted range of values for the initial $T_h$, seem to indicate a scenario for the final stages of quark drop evaporation which is rather different from the one proposed in [22].

We will first present the results obtained for the following combination of values for the free parameters in our model: we consider an initial quark drop of radius $R_s = 10^5$ fm, surrounded by a hadron plasma at temperature $\hat{T}_h = 0.99$, to which is associated a phase interface with surface tension parameter $\sigma_0 = \sigma/T_c^3 = 1$. (We take $T_c = 150$ MeV.) Results with other initial values for $R_s$ and $\hat{T}_h$ show broadly similar hydrodynamical behaviour and the choice for the value of $\hat{T}_h$ is related to the fact that the computational times increase greatly for $\hat{T}_h$ closer to 1. We first discuss the situation for $\sigma_0 = 1$ as this corresponds to the value used in our previous computations [6,8] and also allows one to see clearly the main features of the hydrodynamical scenario. Results for lower values of $\sigma_0$ will be discussed later.

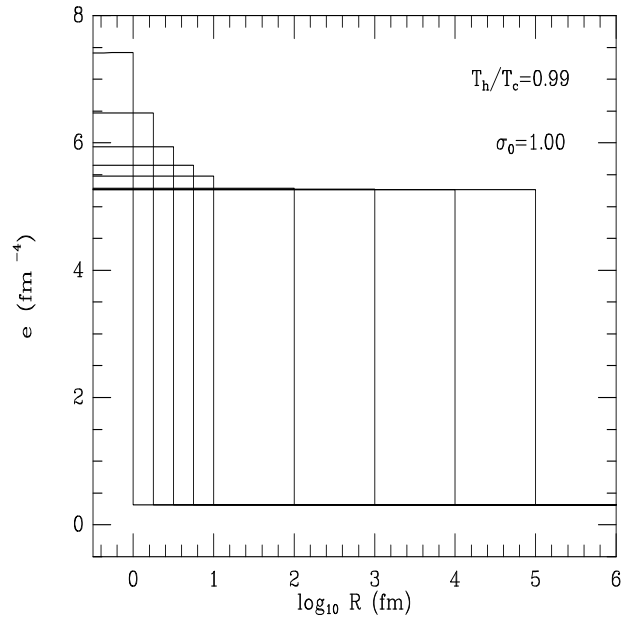

FIG. 5. Time evolution of the profile of the energy density $e$. The hadron phase is on the right of the vertical discontinuity. Note that the similarity solution is preserved until to values of the drop radius of the order of $10^2$ fm.

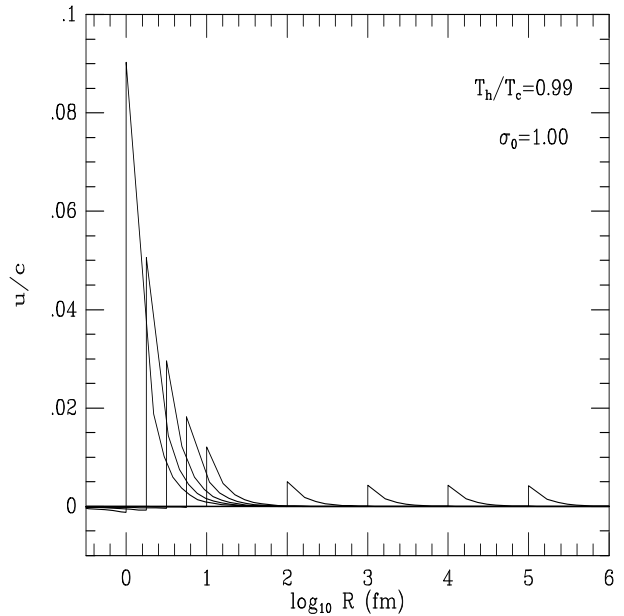

FIG. 6. Time evolution of $u$, the radial component of the fluid four-velocity in the Eulerian frame.

During the initial stages of the drop evaporation, the evolution of the hydrodynamical variables in both phases is in very good agreement with the similarity solution. As the evaporation proceeds and the drop reaches dimensions of the order of $R_s \approx 10^2$ fm, the hydrodynamics starts to deviate significantly from the similarity solution (the evolution is no longer scale-free). The surface tension starts to cause the compression inside the shrinking drop to increase, raising the temperature jump between the two phases which has been constant up to this stage. The value of $\hat{T}_h^-$ is not significantly changed by this, however. This deviation from the similarity solution is at the origin of a run-away mechanism which amplifies the temperature difference between the two sides of the interface and increases the quark evaporation rate. Although the area of the drop surface is reduced, the velocity of the evaporating matter has become larger and this preserves a significant outward flux away from the surface. As a consequence, the contraction is accelerated and the drop experiences an increasingly rapid evaporation which ends with its complete disappearance (see Figs. 5-7)

In this scenario then, the final stages of the evaporation are not regulated by the expansion of the universe, as suggested by [22], but rather by the run-away mechanism which allows the temperature jump across the phase interface to increase. One way of understanding this process is by taking into account the larger outward flow produced by a higher evaporation rate from the drop surface. This can, in turn, increase the efficiency in the removal of the latent heat released, preventing the temperature in the hadron phase rising above $T_c$, while allowing the temperature in the quark phase to grow above $T_c$. The hydrodynamical behaviour described above has been observed for all of the allowed values of the hadron temperature and although the use of lower initial values for $\hat{T}_h$ (to which correspond larger temperature jumps between the two phases) produces a more dramatic and rapid evolution for the shrinking quark drop, all of the results obtained share the above features. A physical limit to the values of $\hat{T}_h$ is given by $\hat{T}_h = 0.61$ which corresponds to a deflagration front moving at the sound speed with respect to the medium behind (i.e. a Chapman-Jouguet deflagration).

The deviation of the hydrodynamical solution away from the self-similar one can be seen as a consequence of the existence of a surface tension $\sigma$. When the drop dimensions become comparable with the characteristic length $l \approx \sigma/(e_q + p_q)$, the drop evolution can no longer be considered as scale-free and surface effects start to become dynamically relevant. When $\sigma_0$ is taken to be exactly zero, the similarity solution continues to hold all the way down to the complete disappearance of the drop. A numerical computation of this situation has been performed and the self-similar solution was preserved with a precision close to the sixth decimal place. If $\sigma_0 = 0.02$, as suggested by recent lattice-gauge simulations [23], the breaking of the similarity solution starts to become relevant later, for drop dimensions $R_s \approx 10$ fm, with the subsequent evolution following the lines described for the case $\sigma_0 = 1$. A main difference between these two situations is given by the magnitude of the relative change in the hydrodynamical variables as the contraction proceeds. Bearing in mind that the sooner the self-similar solution is broken, the more the variables can grow away from their initial values, we report in Table 1 the total relative change of the most relevant quantities as calculated between the initial value (*i.e.* the value at the time the computation is started) and the final value (obtained when $R_s \approx 1$ fm), for different values of $\sigma_0$. A graphical representation of the behaviour of the radial component of the interface four-velocity in the Eulerian frame $u_s$, and of $\hat{T}_{q,h}$ is also given in Figs. 8 and 9. We note that the approximation of treating the phase interface as an exact discontinuity certainly breaks down before the drop radius reaches 1 fm and so our results for the smallest radii should only be treated as indicative.

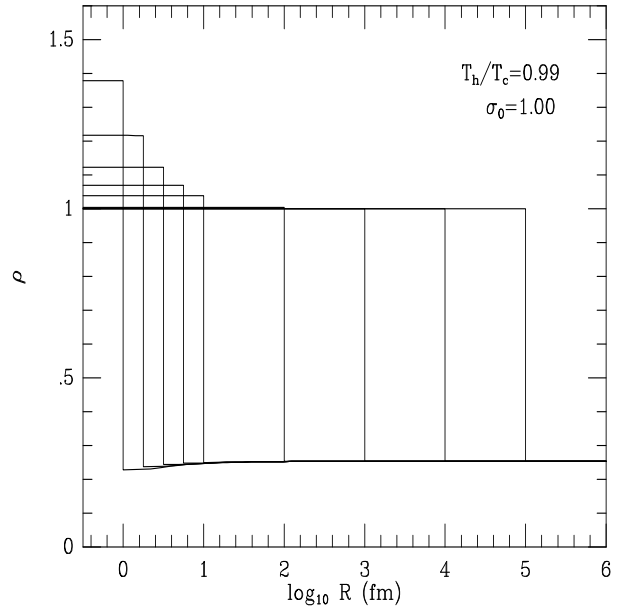

FIG. 7. Time evolution of the compression factor $\rho$. Note that during the contraction of the drop from 10 fm to 1 fm, a depression appears in the hadron phase. This depression will later be compensated by a rarefaction wave.

| $Q$ | $\sigma_0 = 1.00$ | $\sigma_0 = 0.10$ | $\sigma_0 = 0.02$ | $\sigma_0 = 0.00$ |
|---|---|---|---|---|
| $u_s$ | 12.46 | 1.22 | 0.31 | $2.10 \times 10^{-3}$ |
| $e_q$ | 0.41 | $4.01 \times 10^{-2}$ | $8.07 \times 10^{-3}$ | $< 10^{-6}$ |
| $\rho_q$ | 0.49 | $2.99 \times 10^{-2}$ | $7.90 \times 10^{-3}$ | $< 10^{-6}$ |
| $\hat{T}_h$ | $-1.21 \times 10^{-3}$ | $-1.01 \times 10^{-4}$ | $< 10^{-6}$ | $< 10^{-6}$ |
| $\hat{T}_q$ | 0.11 | $9.62 \times 10^{-3}$ | $2.30 \times 10^{-3}$ | $< 10^{-6}$ |

TABLE I. The relative change ($\Delta Q/Q_0$) of the hydrodynamical variables for different values of the surface tension $\sigma_0$. The relative variation is evaluated between the final and the initial values of the relevant variable which is shown in the first column. Here $u_s$ is the radial component of the interface four-velocity as measured by an Eulerian observer. All variations have been calculated for an initial $\hat{T}_h = 0.99$.

One way of looking at the results in Table 1 is that of considering the self-similar solution as a perfect balance between competing effects. As long as the similarity holds, there is a dynamical equilibrium by means of which the evaporation flux reduces the dimensions of the quark drop without increasing the compression or the energy density inside it. The closest analogue of this process could be that of a "leaky" balloon which, although shrinking, maintains the same shape and internal compression by means of ejecting internal material. When the self-similar solution is broken, the balance is lost and the run-away mechanism sets in. Although these very final stages in the life of the quark drop represent only a small fraction of its whole evolution, they can be extremely important since a considerable change in the fluid variables might occur then.

A comment should be made on the results obtained for low initial values of the hadron temperature (*i.e.* $\hat{T}_h \leq 0.80$). Although these cases probably have no cosmological relevance, as they require a strongly supercooled quark plasma, they provide important information about the hydrodynamics of weak deflagration fronts near the stability threshold given by the Chapman-Jouguet point. In the present scenario this occurs for an initial value $\hat{T}_h = 0.61$ and has turned out to be the limit for the numerical solution. While a detailed analysis would go beyond the scope of this paper, it is interesting to note, when looking at the sequence of the different solutions in the range of low $\hat{T}_h$, that the hydrodynamical evolution becomes less stable and regular as values of $\hat{T}_h$ closer to the Chapman-Jouguet point are used. It is particularly interesting to note the behaviour of $u_s$, which is larger for lower initial values of $\hat{T}_h$. In this case, the presence of oscillating modes is very clear and these can perhaps be related to the instabilities which are known to appear from the linear perturbation anal-ysis of two-dimensional deflagration fronts [24,25]. At the Chapman-Jouguet point, the solution is extremely unstable and is almost immediately dominated by instability modes with very large amplitude and very short dynamical time scale. As stated before, it was not possible to calculate any numerical solution beyond this limit. Figs. 10 and 11 show the behaviour of $u_s$ for $\hat{T}_h = 0.80$ and $\hat{T}_h = 0.61$. It is important to point out that, despite the dramatic oscillatory behaviour seen for $u_s$ at low $\hat{T}_h$, the solution for the other quantities remains comparatively smooth and regular up until the Chapman-Jouguet point, beyond which stability is lost completely. This sort of behaviour seems to be characteristic of deflagrations and is related to phenomena observed in laboratory experiments [26]. The complete loss of stability past the Chapman-Jouguet point (i.e. for strong deflagrations), may be related to the fact that in this case the front ceases to be in mutual causal connection with the fluid behind (relative to which the front is supersonic).

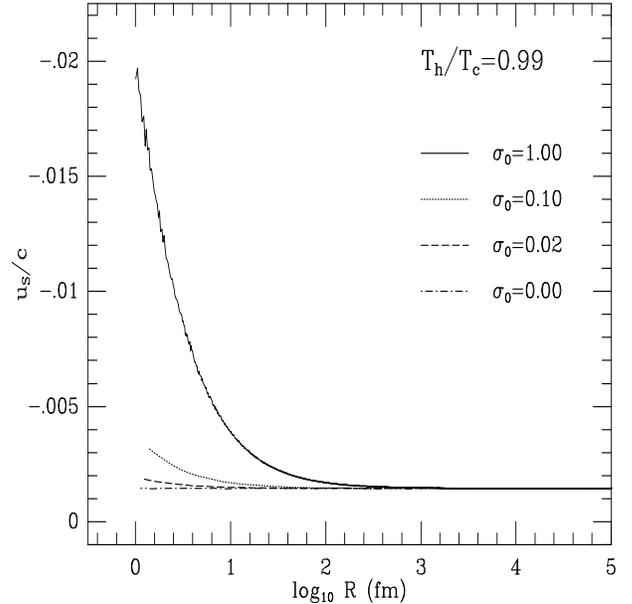

FIG. 8. Behaviour of $u_s$ for different values of the surface tension parameter $\sigma_0$. Note that the deviation away from the self-similar solution becomes extremely small for smaller values of $\sigma_0$.

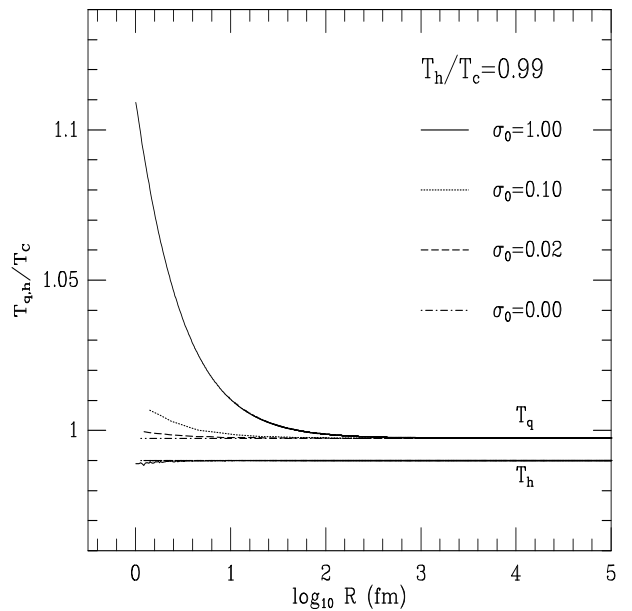

FIG. 9. Behaviour of the temperature in the two phases $\hat{T}_{h,q}$ for different values of the surface tension parameter $\sigma_0$. Note that the deviation away from the self-similar solution becomes extremely small for smaller values of $\sigma_0$.

Much of the astrophysical interest in a first order quark–hadron phase transition concerns baryon number segregation within the shrinking quark drop and the consequent generation of inhomogeneities in the baryon number density. The occurrence of baryon number segregation is related to the fact that in chemical equilibrium, baryon number in the high temperature phase is carried by almost massless quarks, while in the low temperature phase it is carried by nucleons, whose rest mass $m \gg T_C$ and whose number density is limited by a factor $\exp(-m/T_C)$. The degree of inhomogeneity might then be further increased during evaporation of the quark drops. Several studies have been carried out (see [27] for a review) in order to calculate the amplitude of the inhomogeneities which could be produced, but the lack of a detailed knowledge of the micro-physical processes occurring at the interface (where suppression mechanisms could, in principle, reduce the flow of baryon number from the high temperature phase), and the difficulty in estimating the typical scale at which the hadronic bubbles meet, make these calculations extremely tentative.

With the present calculation, in which we neglect the effects of long-range energy and momentum transfer, we have aimed to establish whether, by means of purely hydrodynamical mechanisms, the contraction could significantly aid baryon concentration. With regard to this, a key result is the demonstration of the existence of self-similar solutions which tend to keep constant the values of the compression factor in the two phases. If one makes the (very idealised) assumption that baryon number is carried along exactly with the hydrodynamical flow and limits attention to material which was within the disconnected quark regions at the time when chemical equilibrium was broken, then the baryon number density will be found to follow the same behaviour as the compression factor. As can be seen from Fig. 7, no strong enhancement of the compression factor seems to appear. For a standard quark drop of $\sigma_0 = 1$ and initial $\hat{T}_h = 0.99$, the relative increase of the compression at the end of the contraction is $\Delta\rho/\rho_o \sim 0.49$ (see Table 1). Slightly larger compressions have been found for lower values of $\hat{T}_h$ (e.g. $\Delta\rho/\rho_o = 0.55$ for $\hat{T}_h = 0.80$), but such low values of the temperature seem unlikely to be relevant in the cosmological situation.

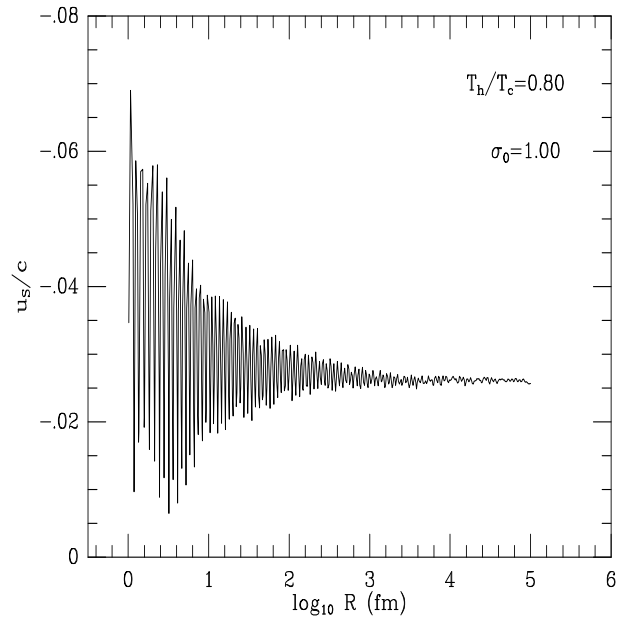

FIG. 10. Time evolution of the radial component of the interface four-velocity in the Eulerian frame for large supercooling. This should be compared with Fig. 8, where $T_h/T_C = 0.99$.

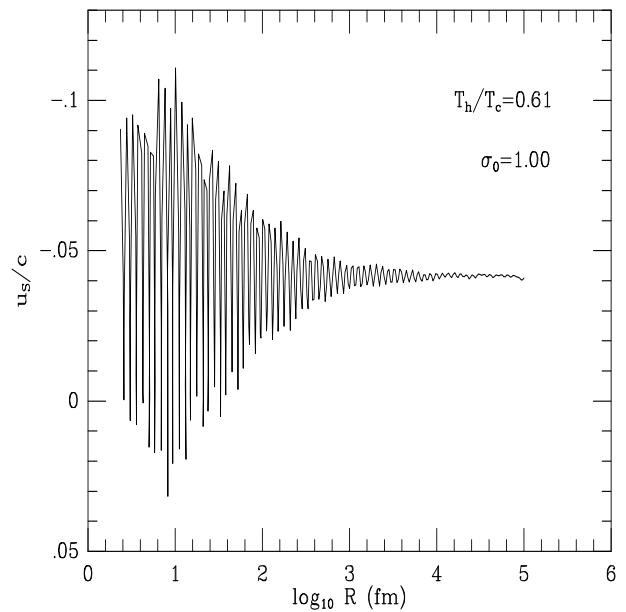

FIG. 11. Time evolution of the radial component of the interface four-velocity in the Eulerian frame for large supercooling. This should be compared with Fig. 8, where $T_h/T_C = 0.99$.

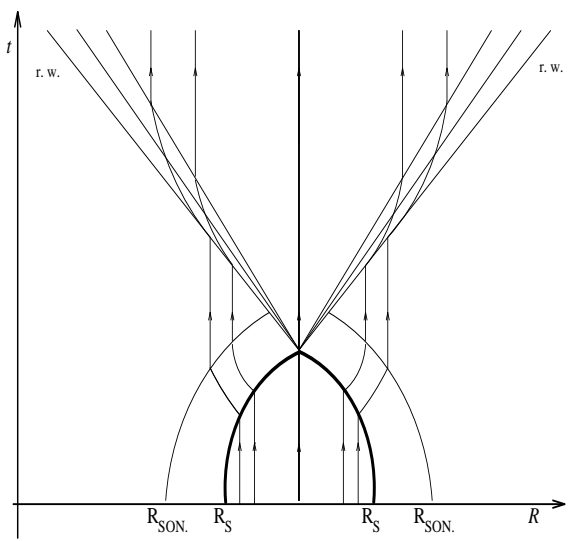

FIG. 12. Space-time diagram of the final stages of the drop evaporation. Here $R_{SON}$ marks the sonic radius, while the thick solid line is the world-line of the phase interface. The oblique straight lines emanating from the point of disappearance of the drop, represent the rarefaction wave (r.w.) and the lines marked with arrows are fluid flowlines.

Our conclusion is that hydrodynamical compression on its own can do little to enhance baryon number concentration. If there is to be a concentration of the magnitude discussed in the literature [11,4], then this must result mainly from the role of the long-range energy transfer or from suppression of the flux of baryon number across the interface. However, it is worth noticing that the final disappearance of the quark drop will be followed by the emission of a rarefaction wave as shown schematically in the space-time diagram in Fig. 12. This will act in the direction of reducing any overdensity which could have formed and its effect needs to be properly taken into account. A more extended treatment is now in progress involving the solution of the relativistic radiative transfer problem together with the solution of the deflagration hydrodynamics for a contracting drop. (The equivalent calculation was carried out in [7,8] for an expanding bubble.)

## VI. CONCLUSION

We have presented the relativistic hydrodynamics of a contracting quark drop during the cosmological quark–hadron phase transition in the absence of long-range energy and momentum transfer. We have shown the existence of a self-similar solution which governs the hydrodynamics of a contracting quark region in the absence of intrinsic length scales. Numerical solution of the hydrodynamical equations for a representative range of values within the allowed parameter space, has led to a new and consistent scenario for the final stages of the phase transition in the absence of long-range energy and momentum transfer. An isolated quark drop with an initial radius large enough so that surface effects can be neglected, has been numerically followed during the contraction. Initially the evaporation is governed by the similarity solution which has the peculiarity of preserving the values of the energy density and of the compression factors in both phases. When the bubble radius becomes sufficiently small (i.e. $R_s \approx \sigma/(e_q + p_q)$), surface effects start to become relevant and the self-similar solution is then broken. The drop then experiences an increasingly rapid evaporation, eventually ending with the emission of a rarefaction wave when the drop finally disappears. This hydrodynamical behaviour has been observed for values of the temperature in the hadron phase ranging between $\hat{T}_h = 0.999$ and $\hat{T}_h = 0.61$, which is the stability threshold for a weak deflagration front (i.e. the Chapman-Jouguet point).

The simulations performed seem to indicate that in the absence of long-range energy and momentum transfer and with baryon number being entirely carried along with the hydrodynamical flux, no relevant baryon number concentration appears above that indicated from considerations of chemical equilibrium in the middle part of the transition. This study provides a hydrodynamical background for a more complete treatment in which long-range energy and momentum transfer via electromagnetically interacting particles will be taken into account.

## ACKNOWLEDGMENTS

LR gratefully acknowledges hospitality from the Astrophysics Group at the University of Oxford where part of this work was done. Financial support for this research has been provided by the Italian Ministero dell'Università e della Ricerca Scientifica e Tecnologica.